\documentclass[epj]{svjour}
\usepackage{graphics}
\usepackage{epstopdf}
\usepackage{subfigure}
\usepackage{float}

\usepackage[T1]{fontenc}
\usepackage{graphicx}
\usepackage{amsmath,amssymb,slashed,bbm,xcolor}
\DeclareMathAlphabet{\boldmathe}{T1}{cmr}{bx}{it}

\newcommand{\n}[1]{\label{#1}}

\renewcommand{\vec}[1]{{\bf #1}}

\def\beq{\begin{eqnarray}}
\def\eeq{\end{eqnarray}}
\def\ln{\,\mbox{ln}\,}

\def\be{\beta}

\def\ga{\gamma}
\def\de{\delta}

\def\om{\omega}

\def\Ga{\Gamma}

\def\Om{\Omega}

\def\B1{ Bianchi-I }

\begin{document}
\sloppy

\title{Cosmic anisotropy with Reduced Relativistic Gas}


\author{Simpliciano Castardelli dos Reis\inst{1}
\thanks{\emph{E-mail address:} simplim15@hotmail.com},
Ilya L. Shapiro\inst{1,2,3}\thanks{\emph{E-mail address:} 
shapiro@fisica.ufjf.br}
%
}                     
%
%
\institute{Departamento de F\'{\i}sica, ICE, Universidade Federal de Juiz de Fora \\
Campus Universit\'{a}rio - Juiz de Fora, 36036-330, MG, Brazil
\and \
Tomsk State Pedagogical University, Tomsk, 634041, Russia
\and \
Tomsk State University, Tomsk, 634050, Russia}

\date{Received: date / Revised version: date}
%

\abstract{The dynamics of cosmological anisotropies is investigated 
for Bianchi type I universe filled by a relativistic matter represented
by the reduced relativistic gas model (RRG), with equation of state
interpolating between radiation and matter. Previously it was shown
that the interpolation is observed in the background cosmological
solutions for homogeneous and isotropic universe and also for
the linear cosmological perturbations. We extend the application
of RRG to the Bianchi type I anisotropic model and find that the
solutions evolve to the isotropic universe with the pressureless
matter contents.
} 
\authorrunning{Simpliciano Castardelli dos Reis, I. Shapiro}
\titlerunning{Cosmic anisotropy with Reduced Relativistic Gas}
\maketitle

\section{Introduction}
\label{Sect1}

The standard cosmological model describes a universe with
 homogeneous and isotropic geometry. The matter contents is
 described by a set of cosmic fluids satisfying some equations of
 state  (EoS). The inhomogeneities are allowed only in the form
 of small perturbations, which define most of the observables
 which are used to define most of relevant observables. We know
 that the spectrum of CMB, Large Scale Structure, BAO and other
 observations demonstrate the correctness of this description,
 that is the dynamics of perturbations proves that the expanding
 universe is very close to homogeneity and isotropy of expanding
 universe at the sufficiently large scale. The question is  whether
 universe was ``born'' isotropic and homogeneous or it became
 such due to some internal mechanism at the early stage of its
 evolution.
The complete analytical description of the possible anisotropies
and non-homogeneities of the early universe is impossible. Therefore
the standard approach is to assume certain symmetry of the metric
tensor.
For instance, homogeneity and isotropy are possible symmetries.
In formulating a more general metrics one of the possibilities is to
consider an anisotropic but homogeneous space-time. The pioneer
work \cite{Kantowski:1966te} explored the case of the metric
anisotropic in two space directions with a given group of symmetries,
in the universe filled by dust, while \cite{Ellis:1966ta} dealt with the
special cases of locally rotationally symmetric and shear-free dust.

The homogeneous models can be grouped by the possible space
symmetries given by the Bianchi classification, which is based
on the Lie algebras satisfied by the Killing vectors or, equivalently,
the structure constants of the hypersurface's tetrad system
\cite{landau1987}. The first possible space of this classification,
called Bianchi type I, has three Killing vectors corresponding to the
three spatial translations. Due to the simplicity of the Bianchi-I metric,
it was extensively used in anisotropic  cosmological models, including
the Kasner vacuum solution \cite{Kasner:1921zz}. More sophisticated
cosmological models based on other types of Bianchi classification, are
possible. One can mention, for instance, the renowed works on the
Bianchi type IX models by Belinskii, Khalatnikov and Lifshitz \cite{Belinsky:1970ew,Belinsky:1982pk}, and the Mixmaster universe
model by  Misner \cite{Misner:1967uu,Misner:1969hg}, which shows
a chaotic behaviour.

One of the main questions concerning anisotropic cosmological
models is whether the universe could be anisotropic in the early
epoch and evolve to be isotropic? It is certainly interesting to
identify a mechanism which could be responsible by such an
isotropization. It is highly desirable to have a maximally simple
description of such a universe, such that further analysis of the
perturbations could provide an observational evidence of
isotropization. Some of mechanisms of this kind consist from the
analysis of the asymptotic behaviour of solutions with isotropic
classical fluids \cite{Jacobs:1968zz,Ellis:1968vb}
(see also \cite{Ellis2012}), viscous  and anisotropic stress tensor
\cite{hervik2007}, primordial magnetic field
\cite{Jacobs:1969qca,Thorne:1967zz} and quantum effects in
primordial universe \cite{Lukash:1976kr,Hu:1978zd}.

In what follows we concentrate on the \B1 model. From the
mentioned references we know that the speed of isotropization
of the metric may depend of the EoS of the contents of the universe
and, in particular, is different for matter or radiation.
The situation is qualitatively similar to the dynamics of metric
and density perturbations on the isotropic background, but we
do not need to treat anisotropies as small perturbations.

Usually, the EoS is assumed to be a linear relation between pressure
and energy density, $p = \om\rho$, with a constant $\om$. The value
of $\om$ corresponds to the type of a fluid. For example, $\om = -1$
means cosmological constant, $\om=0$ dust and $\om=1/3$
radiation. According to the recent data (see, e.g., \cite{Riess:1998cb}
and \cite{Bergstrom:2000pn}) the present-day universe is dominated
by non-luminous sources, such as Dark Matter (DM) and Dark Energy.
It is most likely that the DM is a gas of weakly interacting massive
particles, while the main candidate to be Dark Energy is the
cosmological constant. The observational data show that most of its
history universe was very isotropic, and therefore the isotropization
should occur very early. Since in the past the universe was much hotter
than now, the contribution of the cosmological constant to the overall
energy density balance at the epoch of isotropization was very small
\cite{Bludman}. At the same time, regardless the mass and warmness
of the DM particles are unknown, the DM is supposed to be very hot
in the early universe and then to became relatively cold at the later stage.
Therefore it makes sense to explore the isotropization mechanism for
the case of a universe filled by baryonic and dark matter, which are
hot in the early universe and dust-like in the present epoch. The
simplest appropriate description for the particles in a very early
universe is the ideal relativistic gas of massive particles. Perhaps the
most useful representation of such a gas is through the Reduced
Relativistic  Gas model (RRG), which provides a simplified
approximation to the Maxwell distribution.

The EoS of RRG was originally invented by A.D. Sakharov
in the famous 1966 paper \cite{Sakharov:1966aja}, to interpolate between
radiation and dust regimes. In this work the interpolating EoS has
been used for the first derivation of the CMB spectrum, but the
details of how to obtain the EoS of the model were not given.
More recently RRG model was reinvented by our group in
Refs.~\cite{FlaFlu,Fabris:2008qg}. The main advantage of
this model includes the fact that the solutions for the background
cosmology can be obtained in a closed, analytic form for a wide
class of models including RRG and other fluids
\cite{Medeiros:2012ud}, while the EoS is very close to the one of the
relativistic gas of ideal particles \cite{FlaFlu}. Consequently RRG has
been used for a simplified evaluation of the bounds of warmness of DM
\cite{Fabris:2008qg,Hipolito-Ricaldi:2017kar}, description of
energy exchange between matter and radiation and for an overall
rough estimate for the cosmological observables in the model with
the running cosmological constant \cite{Fabris:2011am}.

In the present work we apply RRG to describe the isotropization of
the universe in the transition period when the matter contents of the
universe is in the transition from the radiation to the dust EoS. We
will follow the classical works \cite{Jacobs:1968zz,Jacobs:1969qca},
but instead of dealing with radiation and dust cases separately,
consider the RRG fluid which interpolates smoothly between the
two regimes.

The paper is organized as follows. In Sec. \ref{Sect2} we present a new
derivation of the EoS of the RRG \cite{Sakharov:1966aja}. This new
derivation is instructive and more formal than the previous one in
\cite{FlaFlu}. In Sec. \ref{Sect3} we formulate the equations
describing the dynamics of Bianchi type I model in the universe
filled by RRG. Sec. \ref{Sect4} describes the simplest approximation
for solving these equations. In particular it is shown that the
previously known radiation and dust cases represent the limiting
cases of the new system of equations. The solution in the general
case of RRG can be possible only by means of numerical methods,
as described in Sec. \ref{Sect5}. Finally, in Sec. \ref{Sect6} we draw our
conclusions and describe the perspectives for the further work.

\section{Reduced relativistic gas: equation of state}
\label{Sect2}

Let us consider the EoS for the RRG model in a way different from
\cite{FlaFlu}.
The model describes ideal relativistic gas of massive identical particles.
The main simplification compared to the J$\ddot{\rm u}$ttner model
\cite{Juttner} (see also the book \cite{Pauli})
is that within RRG particles have identical kinetic
energies. This assumption make the EoS very simple and, in particular,
provides great simplification in cosmology, both at the background
and perturbations level \cite{Sakharov:1966aja}. At the same time,
the difference with the EoS of the J$\ddot{\rm u}$ttner model,
derived on the basis of Maxwell distribution does not exceed $2.5\%$
\cite{FlaFlu}.
For the cosmological applications, since J$\ddot{\rm u}$ttner model
and, in general, an ideal gas of identical particles, is certainly just
an approximation, the RRG is perfectly justified and useful model.

The derivation of EoS in \cite{FlaFlu} is very simple, one can say
it is at at the high-school level. Let us present a little bit more formal
scheme of deriving this equation in the flat Minkowski metric,
which enables one, in  principle, to evaluate the difference with the
J$\ddot{\rm u}$ttner model analytically.

The number of particles $N$ is evaluated on a three-dimensional
space-like hypersurface with the normal vector $n^\mu$, with the
hypersurface element area $d\sigma$. The general expression for a
non-degenerate gas composted of identical particles is \cite{Hakim2011}
\beq
N = \int d\sigma\,d^{4}p\,\,
n_{\mu} p^{\mu}\,f(x,p)\,\delta(p^2-m^2),
\label{equationN}
\eeq
 where $p^2=(p^0)^2-\delta_{ij}\,p^i\,p^j$.   The distribution
 function $f(x,p)$ depends of space-time coordinates and momenta,
 denoted by $x$ and $p$.
 Taking the integral over $dp_{0}$ and using the properties of the
 delta function, we get
\beq
N= \int d\sigma\,\frac{d^{3}p}{p^0}\,n_{\mu}\,p^{\mu}\,f(x,p).
\eeq
For the constant time hypersurface $n^{\mu}=\de^\mu_0$
and $d\sigma=d^{3}x$ we arrive at the expression
\beq
N=\int d^{3}x\,d^{3}p\,\,f(x,p).
\label{equationNconstantTime}
\eeq
 The RRG corresponds to the \textit{ansatz} for  for distribution
 function,
\beq
f(x,p) \,=\, C\,\delta(E-E_{0}),
\eeq
 where $C$ is a normalization constant,
 $E=p^0=\sqrt{{\vec p}^2+m^2}$ and $E_{0}$
 is a constant energy of a gas particle. Using the expression for
distribution function in (\ref{equationNconstantTime}), one can
easily obtain
\beq
N \,= C\int d^3x\,d\Om\,dE\,\,E\,\sqrt{E^2-m^2}\,\de(E-E_{0}),
\eeq
where $d\Om$ is the solid angle element. From the last expression,
one can determine the constant C, leading to the final form of the
distribution function,
\beq
f\,=\,\frac{n}{4\pi E_0 \,\sqrt{E_{0}^2-m^2}}\,\delta(E-E_{0}),
\label{distrubutionfunction}
\eeq
Here $n=N/V$ is the concentration (number of particles per volume)
of the gas. The expression for the energy-momentum tensor is
\cite{landau1987,Hakim2011} (see also brief derivation in the 
Appendix)
\beq
T^{\mu\nu}
\,=\, \int\,d^{3}p\,\,\frac{p^{\mu}p^{\nu}}{p^{0}}\,f(x,p).
\label{energymomentumtensor}
\eeq

In the reference frame of an observer with four-velocity $u^{\mu}$
the projection of the energy-momentum tensor onto the hypersurface
with normal vector $u^{\mu}$ leads to the energy density
$\rho$ and pressure $p$. According to Ref.~\cite{Ellis2012},
\beq
\rho &=& u_{\mu}u_{\nu} T^{\mu\,\nu},
\qquad
p \,=\, -\frac{1}{3}\,h_{\mu\nu}\,T^{\mu\,\nu},
\label{projections}
\eeq
where $\,h_{\mu\nu}= \eta_{\mu\nu} - u_{\mu} u_{\nu}$.
In case of a comoving reference frame, in which observer has the
four-velocity $u^{\mu}=\delta^{\mu}_{0}$ with the distribution
function  (\ref{distrubutionfunction}), the expressions
(\ref{projections}) become
\beq
\rho = n E_{0}
\quad
\mbox{and}
\quad
p = \frac{n(E_0^2-m^2)}{3E_{0}}.
\label{pressure}
\eeq
Defining the rest energy density $\,\rho_d=nm$, the pressure and
energy density are related by the expression
\beq
p &=& \frac{\rho}{3}\,\Big(1-\frac{\rho_d^2}{\rho^2}\Big),
\label{rrgsateequation}
\eeq
which is nothing else but the EoS of the RRG model
\cite{Sakharov:1966aja,FlaFlu}. It is easy to see that this EoS
interpolates between radiation, $p \sim \rho/3$, at high energies,
when $\rho^2 \gg \rho_d^2$, and dust $p \sim 0$, at low
energies, when $\rho^2 \approx \rho_d^2$.

\section{Bianchi-I type cosmology with RRG}
\label{Sect3}
Consider the anisotropic cosmology with the RRG fluid. Our starting
 point will be the space of Bianchi-I type, with the metric of the form
 \cite{landau1987},
\beq
ds^2 = dt^2 - a^2_{1}(t)\,dx^2-a_{2}^2(t)\,dy^2-a_{3}^2(t)\,dz^2.
\label{lineelement}
\eeq
 A useful parametrization of anisotropic metric was introduced by
 Misner in \cite{Misner:1967uu,Misner:1969hg},
\beq
a_{1/2}(t) = a(t)\,e^{\beta_{+}(t) \pm \sqrt{3}\,\beta_{-}(t)}\,,
\qquad
a_{3}(t) = a(t)\,e^{-2\,\beta_{+}(t)},
\label{MisnerParametrization}
\eeq
where $a(t)$, $\beta_{+}(t)$ and $\beta_{-}(t)$ are unknown
functions of time.
In this parametrization $\sqrt{-g}=a_{1} a_{2} a_{3}= a^3\,$
and the relation between $\,\be_{\pm}$ and $a_{i}$ is
\beq
\beta_{+}
=\frac{1}{6}\,\ln\Big(\,\frac{a_{1}\,a_{2}}{a_{3}^2} \,\Big)
\, \qquad
\beta_{-}
=\frac{1}{2\,\sqrt{3}}\,\ln\Big(\,\frac{a_{1}}{a_{2}} \,\Big).
\eeq
In Ref. \cite{Ellis2012}, within the \textit{1+3} covariant formalism
of a system of time-like geodesic congruence, the change of a
connecting vector between geodesics, expressing the relative distance,
is split in the irreducible parts called \textit{shear}, \textit{vorticity}
and \textit{expansion}. In particular, the functions $\beta_{\pm}$ are
the independent components of a traceless tensor, which represents
the shear. In what follows, we analyse the dynamics of gravitational
field for the metric (\ref{MisnerParametrization}), generated by
Einstein equations. The matter contents of the universe is modelled
by an isotropic RRG, where pressure is assumed to be the same in all
spatial directions.

The energy-momentum tensor is
\beq
T_{\mu\nu} = (\,\rho + p\,)\,u_{\mu}\,u_{\nu}+ p\,g_{\mu\nu}.
\label{energymomentumtensor-1}
\eeq
The conservation equation $\nabla_{\mu}\,T^{\mu\nu}=0$ leads to
\beq
\dot{\rho}+ 3\,H\,(\,\rho+p\,)=0,
\qquad
H=\frac{\dot{a}}{a}\,,
\label{conservationequation}
\eeq
where the dot means derivative with respect to the physical time.
Let us note that the anisotropy of the metric does not affect the last
equation because of the isotropic pressure.
Eq.~(\ref{conservationequation}) can be integrated by using
the EoS of the RRG, yielding the same result as in
the isotropic case \cite{FlaFlu},
\beq
\rho = \sqrt{\rho^2_{1}\,\Big(\,\frac{a_{0}}{a} \,\Big)^6
+\rho^2_{2}\,\Big(\,\frac{a_{0}}{a} \,\Big)^8 },
\label{energydensity-rho}
\eeq
where $\rho_{1}$, $\rho_{2}$ and $a_{0}$ are integration constants.

As usual, one can distinguish two extreme regimes in the solution
(\ref{energydensity-rho}).  In the case $\rho_1 \ll \rho_2$, one meets
the ultra-relativistic case, that is RRG demonstrates radiation-like
behaviour. On the other hand, for $\rho_1 \gg \rho_2$, RRG
behaves like a dust.

Now we are in a position to consider the Einstein equations for the
Bianchi - I metric. According to Ref.~\cite{hervik2007}, the Einstein
tensor, $G_{\mu\nu}$ for the metric (\ref{MisnerParametrization})
assumes the form
\beq
G_{0\,0} &=& 3\,H^2
\,-\, 3\big(\,\dot{\beta}_+^2 + \dot{\beta}_-^2\,\big)
\,,
\nonumber
\\
G_{1\,1} &=&
-3H^2 - 2\dot{H}
\,-\, 3\big(\,\dot{\beta}_+^2 + \dot{\beta}_-^2\,\big)
\,+\nonumber\\
& &+\Big(\frac{d^2}{dt^2}
+3H\frac{d}{dt}\Big)\,\big(\,\beta_{+}
+\sqrt{3}\,\beta_{-}\,\big),\nonumber
\\
G_{2\,2}  &=&
-3H^2 - 2\dot{H} \,-\, 3\big(\,\dot{\beta}_+^2 + \dot{\beta}_-^2\,\big)
\,+\nonumber\\
& &+\Big(\frac{d^2}{dt^2}+3H\frac{d}{dt}\Big)
\,\big(\,\beta_{+}-\sqrt{3}\,\beta_{-}\,\big),\nonumber
\\
G_{3\,3} &=&  - 3H^2 - 2\dot{H}
\,-\, 3\big(\,\dot{\beta}_+^2 + \dot{\beta}_-^2\,\big)
\,-\nonumber\\
& &-2\, \Big(\frac{d^2}{dt^2}+3H\frac{d}{dt}\Big)\,\beta_{+}.
\label{eisnteintensor}
\eeq

The Einstein equations are given by  $G_{\mu\nu}= 8\pi G T_{\mu\nu}$.
For an isotropic $T_{\mu\nu}$ tensor, Einstein equations can be
rewritten such that the pressure of matter does not  enter the equations.
Following \cite{hervik2007}, we define the new quantities
\beq
G_{+} &=& \frac{1}{6}\,
\big(\,G_{11}+G_{22}-2\,G_{33}\,\big),
\nonumber
\\
G_{-} &=& \frac{1}{2\,\sqrt{3}}\,
\big(\,G_{11}-G_{22}\,\big),
\label{einsteinmaismenosdefinition}
\eeq
yielding significant simplifications compared to (\ref{eisnteintensor}),
\beq
G_{\pm} &=& \ddot{\be}_{\pm} + 3\,H\dot{\be}_{\pm}.
\label{einsteinmaismenorcalculated}
\eeq
Einstein equations boil down to\footnote{ 
Eqs.~(\ref{einsteintensormaismenos}) also follows from the variation
of the Einstein-Hilbert action with respect to $\,\be_{\pm}$.}
\beq
G_{+}
&=& \frac{8\,\pi\,G}{6}\,\big(T_{11}+T_{22}-2\,T_{33}\big),
\nonumber
\\
G_{-} &=&
\frac{8\,\pi\,G}{2\sqrt{3}}\,\big(T_{11} - T_{22}\,\big).
\label{einsteintensormaismenos}
\eeq
Furthermore, since we assume isotropic energy-momentum tensor,
$T_{11}=T_{22}=T_{33}$ and
\beq
G_{\pm} = 0.
\label{anisotropicequation}
\eeq
Finally, $00$-component of Einstein equations, together with
Eqs.~(\ref{anisotropicequation}) and
(\ref{einsteinmaismenorcalculated}), yield
\beq
H^2 \,-\, \big( \dot{\be}_{+}^2 + \dot{\be}_{-}^2 \big)
&=& \frac{8\pi G}{3}\,\rho
\label{eisnteinequations1}
\\
\mbox{and } \qquad
3H\,\dot{\beta_{\pm}}+ \ddot{\beta_{\pm}}&=& 0.
\label{eisnteinequations2}
\eeq
 A first integral of (\ref{eisnteinequations2}) can be easily found
 in the form
\beq
\dot{\beta_{\pm}} = \ga_{\pm}\,a^{-3},
\label{firsintegral}
\eeq
where $\ga_{\pm}$ are integration constants. The last result
transforms (\ref{eisnteinequations1}) into an equation for the
conformal factor of isotropic expansion $a(t)$. Defining useful
constants $\Ga$ and $\phi$,
\beq
\ga_{+} &=& \Ga \cos \phi
\,,\qquad
\ga_{-} \,=\, \Ga\sin \phi\,,
\label{angularrelations}
\eeq
we arrive at the generalized form of Friedmann equation for
anisotropic \B1 metric with RRG matter contents,
\beq
H^2\,=\, \frac{\dot{a}^2}{a^2}
&=&
\Ga^2\,\Big(\,\frac{a_{0}}{a}\,\Big)^6
\,+\nonumber\\
& & +\frac{8\pi G}{3}\,\rho_1\Big(\,\frac{a_{0}}{a}\,\Big)^4
\,\sqrt{\Big(\,\frac{a_{0}}{a}\,\Big)^{-2}+b^2}\,,
\label{prefriedmann}
\eeq
where $b = \rho_2/\rho_1$ is the warmness parameter
\cite{Fabris:2008qg}.  The specific new element compared to
isotropic cosmological model is the first term in the {\it r.h.s.}.
This term has the ultrarelativistic $a^{-6}$ scaling feature and
hence it is irrelevant for the late cosmology. At the same time, it
may be quite relevant in the early universe. Due the new term,
caused by anisotropy, the very early universe behaves according to
\beq
a \,\sim t^{1/3}\,,
\label{13}
\eeq
different from the radiation-dominated universe. It is well-known that
the same dynamics of the conformal factor can be achieved in the
isotropic plane universe with an ideal fluid with EoS $p=\rho$. In
order to see this consider the EoS $p=w\rho$, with constant $w$.
Using the conservation law results in $\rho = \rho_0 a^{-3\,(w+1)}$.
By comparing this result to the first term of the \textit{r.h.s} of
(\ref{prefriedmann}), we arrive at $w=1$. A fluid of this kind was
called \textit{stiff matter}, when first introduced by Zel'dovich
\cite{Zeldovich:1972zz}. We have seen that this EoS results from
integrating anisotropies at the early stage of the evolution of the
universe.

The solution of Eqs.~(\ref{firsintegral}) can be expressed as
\beq
\beta_{\pm}(t)-\beta^{0}_{\pm} = \ga_{\pm}W(t),
\qquad
W(t)=\int_{t_0}^t \frac{dt'}{a^3(t')}.
\label{firstintegral2}
\eeq
In this expression $t_0$ correspond to the initial
moment of time and $\beta^{0}_{\pm}$ are integration constants. One
can notice that both $\beta_{\pm}$, with exception of the integration
constants, will lead to a same functional form.
After $W(t)$ is found, the parameters
$\,\ga_{\pm}\,$ determine
$\be_{\pm}$ and consequently the metric components by
Eqs.~(\ref{MisnerParametrization}).

\section{Approximations}
\label{Sect4}

It is easy to present the solution for (\ref{prefriedmann}) and
(\ref{firstintegral2}) in the form of quadratures, however the
integrals are not elementary functions and the qualitative
analysis becomes cumbersome.
Therefore, in order to have better idea about the physical output
of these equations, we split the derivation of the scale factor
dependence into two different considerations. In the present
section we consider three approximations, namely, vacuum,
radiation and dust. In the next section we present the results
of a numerical solution in the general case. The radiation and
dust approximations come from the limits of the RRG EoS
depending on the value of parameter $b$. The approximation
for vacuum will be explained bellow. The considerations in this
section are almost completely non-original and are presented as
to serve as reference for the consequent numerical solutions.
When $a(t)$ is very small, one can keep only the first (stiff
matter of anisotropic origin)  term on the \textit{r.h.s} of
 (\ref{prefriedmann}).  This procedure is equivalent to taking a
 vacuum solution, because in this regime we are disregarding the
 terms coming from the matter contents. Indeed, it is known that
 for the evolution of homogeneous and anisotropic models in the
 vicinity of the singularity the matter contents has no much
 relevance \cite{Lifshitz:1963ps} (see also \cite{Zeldovich:1983cr}).

 The vacuum metric of Bianchi type I  is called Kasner solution.
 Following \cite{hervik2007}, we arrive at
\beq
\frac{\dot{a}^2}{a^2} = \Ga^2\Big(\frac{a_{0}}{a}\Big)^6,
\eeq
 which can be solved in the form
\beq
\Big(\frac{a}{a_{0}}\Big)^3 = 3\Ga (t-t_{0}).
\label{kasner1}
\eeq
Setting $a_{0}=1$, the equations (\ref{firstintegral2}) can be
integrated, yielding
\beq
\beta_{\pm}(t)\,=\,\be^{(0)}_{\pm} \,+\,\frac{\ga_{\pm}}{3\,\Ga}\,\ln\Big(\,\frac{t}{t_{0}}\,\Big).
\label{kasner2}
\eeq
Here $\be^{(0)}_{\pm}$ and $t_{0}$ are integration constants.
From the angular relations (\ref{angularrelations}) the functions
$a_k(t)$ can be presented as
\beq
a_k(t) &=& (3\Ga)^{1/3}\,t^{p_k}\,,\qquad k=1,2,3\,.
\label{kasner3}
\eeq
Parameters $p_k$ can be written down using notations
(\ref{angularrelations}),
\beq
p_{1/2} &=& \frac{1}{3}\,\big(1+ \cos\,\phi \pm \sqrt{3}\,\sin\,\phi\big)\,,
\nonumber
\\
p_{3} &=& \frac{1}{3}\,\big(1- 2\, \cos\,\phi \big)\,.
\label{kasnerparameters1}
\eeq
and the line element as
\beq
ds^2 = dt^2 - \big(3\Ga\big)^{2/3}\,\big[t^{2p_{1}}\,dx^2
- t^{2p_{2}}\,dy^2 - t^{2p_{3}}\,dz^2\,\big],
\label{kasnerelement1}
\eeq
The multiplicative constant can be absorbed into the spatial
coordinates, providing the standard form \cite{landau1987},
\beq
ds^2 = dt^2 - t^{2p_{1}}\,dx^2- t^{2p_{2}}\,dy^2-t^{2p_{3}}\,dz^2,
\label{kasnerelement2}
\eeq
where the parameters $p_k$, $p_{2}$ and $p_{3}$ satisfy the
algebraic constraints
\beq
p_{1}^2+p_{2}^2+ p_{3}^2 = 1 ,\quad
p_{1}+p_{2}+ p_{3} =1.
\label{kasnerrelations}
\eeq
Finally, in the Kasner solution
\beq
a(t) = \big[\,a_{1}(t)\,a_{2}(t)\,a_{3}(t)\,\big]^{1/3}
=t^{\frac{1}{3}\,(p_{1}+p_{2}+p_{3})}=t^{1/3}.
\eeq

The approximations for which the analytic solution can be
easily obtained correspond to the ultra-relativistic,
$b^{-1}\to 0$ or dust, $b\to 0$ regimes. In what follows we consider
these two cases separately. Let us note that the general form of
solutions (\ref{firstintegral2}) remains the same independent on
the approximations for the isotropic energy-momentum tensor.

In the ultra-relativistic case one can perform the expansion up to
the first order in $b^{-1}$ in (\ref{prefriedmann}). Taking
$a_{0}=1$, we arrive at
\beq
\dot{a}^2 \,=\, \frac{\Ga^2}{a^4}
+ \frac{8 \pi G\rho_{1}b}{3a^2}
\,\Big(1+\frac{a^2}{2b^2}\Big).
\label{radiationaprox1}
\eeq
Taking into account the $a^2/b^2$-term in the parenthesis,
this is the Bianchi type I model with radiation, which has initial
density expressed by $\rho_{2}$. This is exactly the classical
result of  \cite{Jacobs:1968zz} for the radiation, but we obtained
it as a limit of the RRG solution.

It proves useful to make a change of variables
\beq
a &=& \frac{1}{\chi_{rad}}\,\sinh\,\xi,
\quad
\kappa_{\rm rad}^2 \,=\, \frac{8\pi G\,\rho_2}{3\Ga^2},
\quad
0\leq\,\xi\,<\infty\,,
\label{implicit1}
\eeq
in Eq.~(\ref{prefriedmann}). This results in the relation
\beq
dt \,=\, \frac{1}{\Ga\kappa_{\rm rad}^3}\,\sinh^2\xi \,d\xi\,.
\eeq
Then Eqs.~(\ref{prefriedmann}) and (\ref{firstintegral2})
become the parametric relations
\beq
\Ga t &=& \frac{1}{4\kappa_{\rm rad}^3}\,(\sinh 2\xi \,-\, 2\xi)\,,
\quad
\be_{\pm} \,=\, \frac{\Ga_{\pm}}{\Ga}\,\ln (\tanh \xi)\,.
\label{implicitequationsrad}
\eeq
From the relation between $a$ and $\xi$ in (\ref{implicit1}),
one can obtain
\beq
\tanh \xi
&=&
\Big(1+ \frac{1}{a^2\,\kappa_{\rm rad}^2}\,\Big)^{-\,\frac12}.
\label{tgh}
\eeq
In case $(a^2\,\kappa_{\rm rad}^2)^{-1}$ is very small, one gets the
relation
\beq
\tanh \xi &=&
1 - \frac{1}{2}\,\frac{1}{a^2\,\kappa_{\rm rad}^2} \,+\, \dots\,.
\eeq
Consequently, due to the (\ref{implicitequationsrad}),
\beq
\beta_{\pm} \,=\, -\,\frac{\ga_{\pm}}{\Ga}\,
\Big[\,\frac{1}{a^2\,\kappa_{\rm rad}^2} \,+\, \dots
\Big].
\eeq
In the radiation approximation, if we disregard the term
$\big(a^2 \kappa_{\rm rad}^2\big)^{-1}$, then $\be_{\pm}$
tend to zero for great values of $a$, and effectively there is
isotropization.

Another way to arrive at the same conclusion is by observing
that when $\xi \to \infty$, we have $\be_{\pm}\to 0$. In the
same limit
\beq
\sinh\xi \sim \cosh\xi \sim \frac{1}{2}\,e^\xi
\eeq
and dominate in the Eq.~(\ref{implicitequationsrad}). Using
(\ref{implicit1}),
\beq
t \,\approx\, \frac{2a^2}{\kappa_{\rm rad}},
\label{time}
\eeq
which yields the standard expression for the isotropic
radiation dominated universe,
\beq
a = \Big(\frac{2\pi G\,\rho_2}{3\Ga^2}\,\Big)^{1/4}
\,\sqrt{t}.
\n{a}
\eeq
This expression means that the role of anisotropy is
negligible for the evolution of the scale factor and hence
we have isotropization.
\vskip 2mm

Let us now consider the limit $a\gg b$, which means a
dust-dominated universe. The solution of the dynamical equations
(\ref{prefriedmann}) and (\ref{firstintegral2}) for dust is simpler
than for the radiation-dominated case \cite{hervik2007} and was
originally obtained in \cite{heckmann1958}. Here we will try
to arrive at the same result by taking the corresponding limit in
the general solution for RRG, which interpolates between radiation
and dust.

The solutions of Eqs.~(\ref{prefriedmann},\ref{firstintegral2})
for $a(t)$ and $\beta_{\pm}(t)$ are given by
\beq
a^3 = \frac{3\Ga}{t_{I}}\,t\,\big(\,t+t_{I}\,\big),
\qquad
\beta_{\pm}
= \frac{\ga_{\pm}}{3\Ga}\,\ln\Big(\frac{t}{t + t_I} \Big).
\label{dustaprox2}
\eeq
Here
\beq
t_{I}=\frac{4}{3\Ga\kappa_{\rm dust}^2}
\qquad \mbox{and} \qquad
\kappa_{\rm dust}^2= \frac{8\pi G \rho_1}{3\Ga^2}
\label{ka}
\eeq
are constants.
The solutions 
(\ref{dustaprox2}) in the
dust-dominated approximation can be considered in two different
asymptotic situations. The first one is $t \ll t_{I}$, which
implicates in the expansions
\beq
a^3 &=& 3\Ga\,\Big[\,\Big(\frac{t}{t_I}\Big)^2 + t\,\Big],
\nonumber
\\
\be_{\pm} &=& \frac{\ga_{\pm}}{3\Ga}\,
\ln\Big[\frac{t}{t_I}\,\Big(\,1+\frac{t}{t_I}+\dots\,\Big)\Big].
\label{dustkasner}
\eeq
Disregarding terms with powers greater than two, the solutions
tend to  the Kasner expressions (\ref{kasner1}) and (\ref{kasner2}).

In the second case $t \gg t_{I}$ one can use the same scheme
as before, but now making expansion in the powers of $t_I/t$.
In this way we obtain the standard solution for the dust, with
$a \sim t^{2/3}$ and $\beta_{\pm}\to 0$.
Following the same logic as in the radiation case, we conclude
that the behaviour in the late times demonstrates isotropization.

\section{Numerical Solution}
\label{Sect5}

Let us consider numerical solution of the dynamical system
of Eqs.~(\ref{prefriedmann}) and (\ref{firstintegral2}) without
assuming high- or low-energy approximations. Exactly as it was
done in the previous section, we consider a simplified model
with one fluid described by RRG and the anisotropy which
enters the general energy balance by means of the stiff matter
energy density. It proves useful to express the solution in terms
of initial values of the relative energy densities parameters
$\Omega_{an}^{(i)}$ and $\Omega_{RRG}^{(i)}$, defined by
\beq
\Omega_{an}
=
\frac{\Ga^2}{H^2},
\quad
\Omega_{RRG}
=
\frac{8\,\pi\,G\,\rho_{1}}{3\,H^2}\,\sqrt{1+b^2}=1-\Omega_{an}.
\label{Omega}
\eeq

The subscript $(i)$ denotes the values of the parameters in the
initial moment of time. Our purpose is to evaluate the isotropization
of the universe starting from the initial moment of time $t=0$, when
$a(0)=a_{i}=1$ and $H(0)=H_{i}$. Therefore, in the initial instant
of time the values are
$\,\Omega_{an}^{(i)}\,$ and $\,\Omega_{RRG}^{(i)}$, corresponding
to $H=H_i$ in (\ref{Omega}).

It proves useful to define the dimensionless time variable
$\tau = H_i t$. In this way we arrive in the equations
\beq
\frac{\dot{a}^2}{a^2}
&=&
\frac{\Omega_{an}^{(i)}}{a^6}
+ \frac{\Omega_{RRG}^{(i)}}{a^4\,\sqrt{1+b^2}}\,\sqrt{a^2+b^2},
\nonumber
\\
\dot{\be}_{\pm}
&=&
\frac{\sqrt{\Omega_{an}^{(i)}}\,\ga_{\pm}}{\Ga a^3}, 
\label{newsystem}
\eeq
where the dots mean derivatives with respect to $\tau$,

The value of $\Omega_{an}(t)$ measures the amount of anisotropy,
such that greater values correspond to higher degree of anisotropy.
As before, $b$ is the warmness parameter of the RRG matter. In the
nowadays universe the value of $\Ga$ is very small implying in a
very small value of $\Om^0_{an}$. The warmness $b$ today is
bounded from above by approximately $0.001$ for the dominating
fluid, namely for the Dark  Matter \cite{Fabris:2011am}. Indeed, in
the early universe when $\Om^0_{an}$ was significant, the warmness
could have a large value. The framework of RRG enables one to see
how the warmness affects the time of isotropization, that is the typical
time of transition from large value of $\Omega_{an}^{(i)}$ to a small
value at the later period.

The second  equation in (\ref{newsystem}) can be expressed via the
angular parameter in (\ref{angularrelations}). This angle becomes
relevant only in the vicinity of the singularity, when the metric can
be approximated by the Kasner solution, and in the subsequent
numerical analysis it will not play much role.

\begin{figure}[H]
\centering
\includegraphics[height= 5.0 cm,width=\columnwidth]{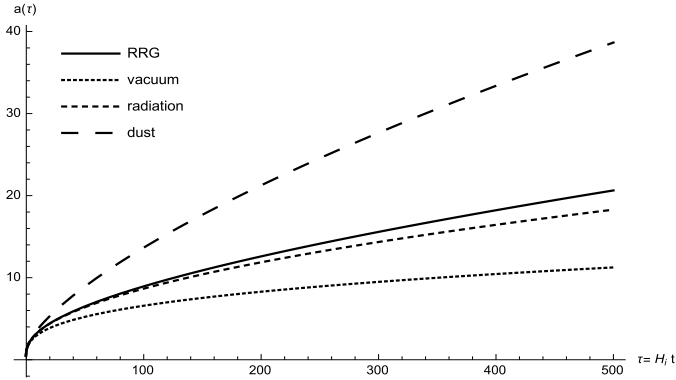}
\caption{Plots of scale factors for the initial values of
$b=10$ and $\Omega_{ani}^{(i)}=0.99$.}
\label{fig1}
\end{figure}

\begin{figure}[H]
\centering
\includegraphics[height= 5.0 cm,width=\columnwidth]{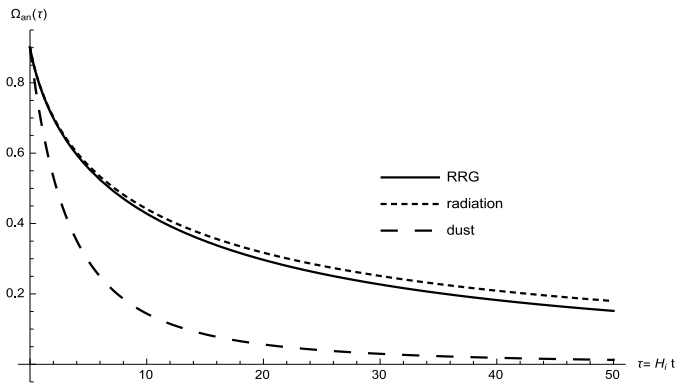}
\caption{Plots of $\Omega_{ani}(\tau)$ for the initial values of
$b=10$ and $\Omega_{ani}^{(i)}=0.99$.}
\label{fig2}
\end{figure}

\begin{figure}[H]
\centering
\includegraphics[height= 5.0 cm,width=\columnwidth]{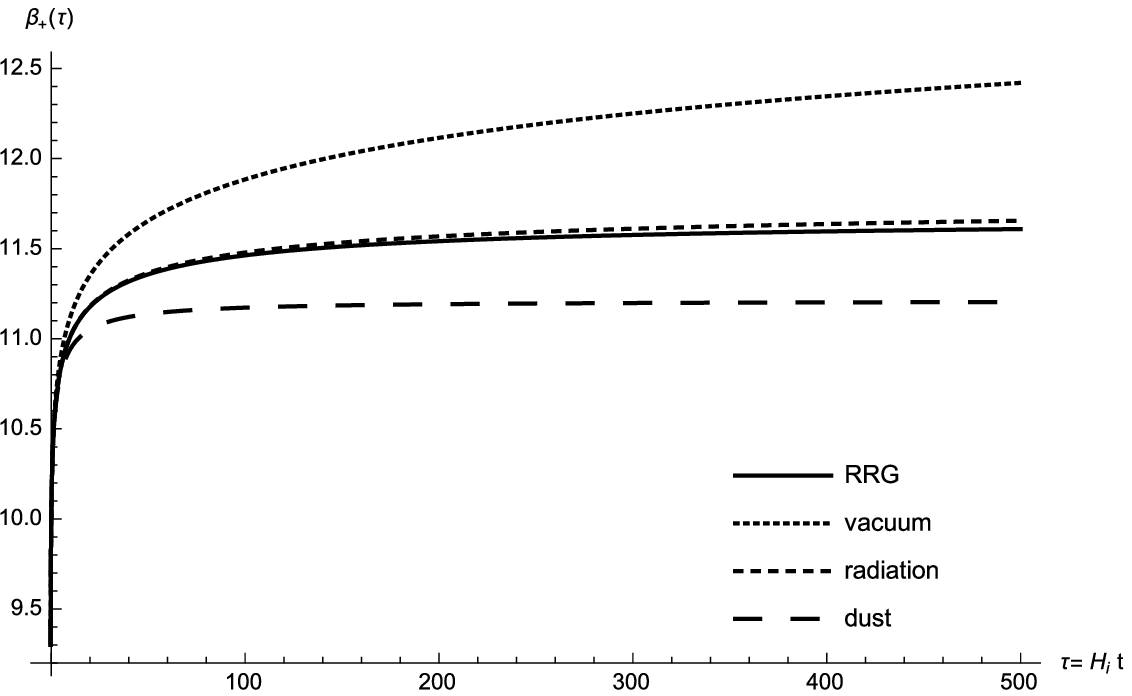}
\caption{Plots of $\be_{+}$ corresponding to the parameters $b=10$ and
$\Omega_{ani}^{(i)}=0.99$, while the initial condition
$\beta_{+}=10$.}
\label{fig3}
\end{figure}

\begin{figure}[H]
\centering
\includegraphics[height= 5.0 cm,width=\columnwidth]{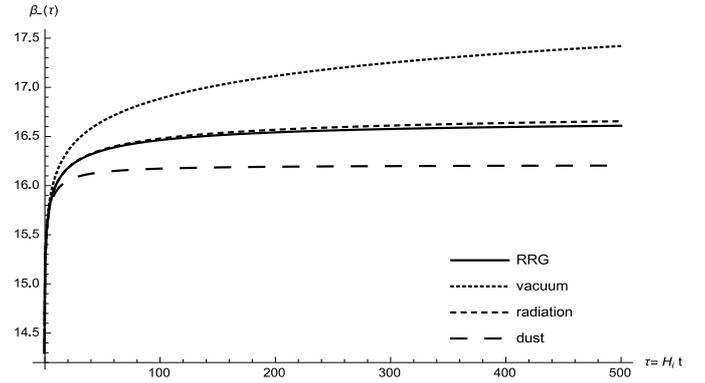}
\caption{Plots of $\be_{-}$ corresponding to the parameters $b=10$ and
$\Omega_{ani}^{(i)}=0.99$, while the initial condition
$\beta_{-}=15$.}
\label{fig4}
\end{figure}

\begin{figure}[H]
\centering
\includegraphics[height= 5.0 cm,width=\columnwidth]{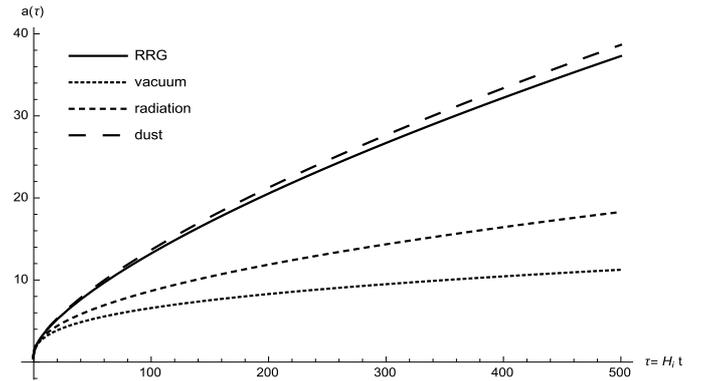}
\caption{Plots of scale factors for the moderate warmness.
Parameters are as follows: $b=0.5$, $\Omega_{ani}^{(i)}=0.99$.}
\label{fig5}
\end{figure}

\begin{figure}[H]
\centering
\includegraphics[height= 5.0 cm,width=\columnwidth]{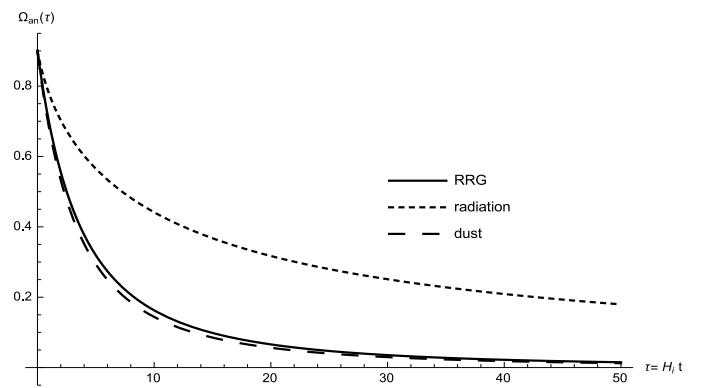}
\caption{Plots of $\Omega_{ani}(\tau)$ for the moderate warmness.
Parameters are as follows: $b=0.5$, $\Omega_{ani}^{(i)}=0.99$.}
\label{fig6}
\end{figure}

\begin{figure}[H]
\centering
\includegraphics[height= 5.0 cm,width=\columnwidth]{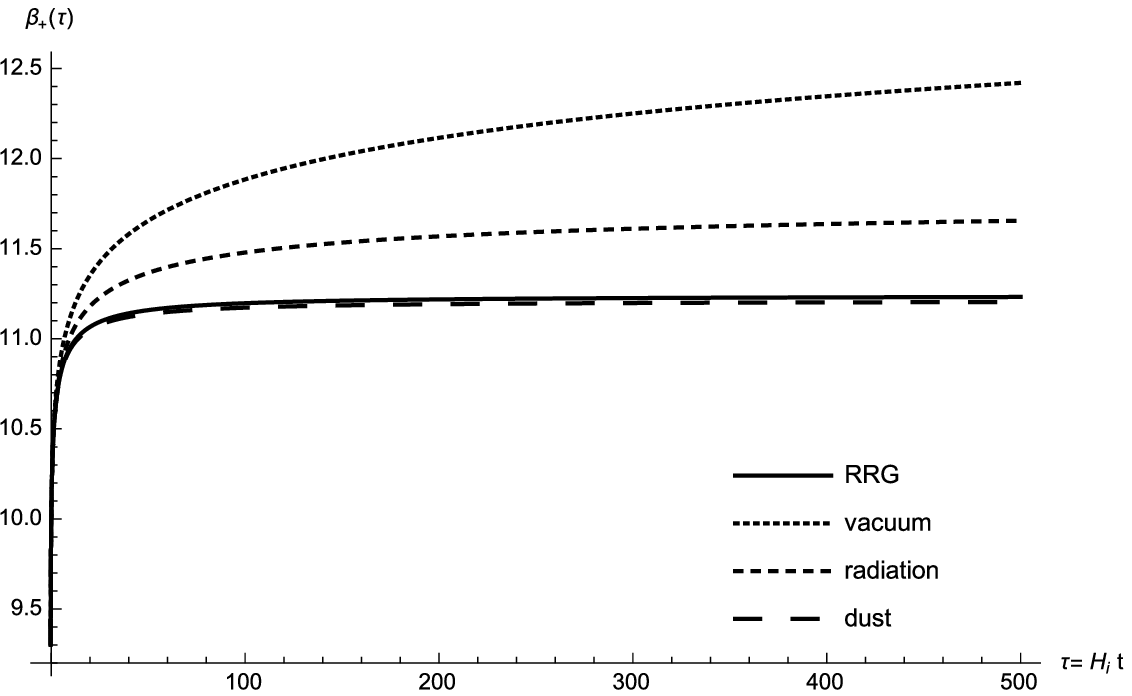}
\caption{Plots of $\be_{+}$ for the moderate warmness.
Parameters are as follows: $b=0.5$, $\Omega_{ani}^{(i)}=0.99$ with 
the initial condition $\beta_{+}=10$.}
\label{fig7}
\end{figure}

\begin{figure}[H]
\centering
\includegraphics[height= 5.0 cm,width=\columnwidth]{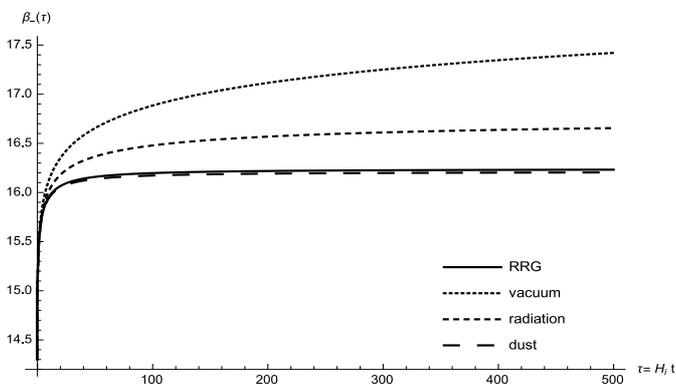}
\caption{Plots of $\be_{-}$ for the moderate warmness.
Parameters are as follows: $b=0.5$, $\Omega_{ani}^{(i)}=0.99$ 
while the initial condition is $\beta_{-}=15$.}
\label{fig8}
\end{figure}

Let us present the numerical solutions for different values of the
warmness parameter $b$ using Mathematica software \cite{M9}.
We used the initial  conditions $a=a_{i}=1$, $\be_{+}=10$,
$\be_{-}=15$, such that
$\Omega_{an}^{(i)}=0.99$ and  $\Omega_{RRG}^{(i)}=0.01$ at
$\tau=0$. In all plots the scale factor and
anisotropy measure $\Omega_{an}$ are compared with the plots
for the cases of vacuum, radiation and dust, by assuming the same
initial values of $\Omega^{(i)}$'s. The Figs.~\ref{fig1} and \ref{fig2} clearly shows
that RRG behaviour tends to Kasner at the early stage of evolution,
and is very close of radiation during some time for both scale factor
and $\Omega_{an}(\tau)$. In the Figs.~\ref{fig3} and \ref{fig4}, the isotropisation can be observed, because
$\be_{+}$ and  $\be_{-}$ tend to constants. It is easy to see that
the isotropization for for RRG occurs faster than for the dust-like
contents, close to the rate in the radiation case.
For smaller warmness, $b=0.5$, one can observe in 
Figs.~\ref{fig5}-\ref{fig8} 
another behaviour, when RRG plot is (quite naturally) close to dust.

The plots presented above show that the RRG is perfectly well
interpolating between radiation and dust regimes, as it should be
expected. The asymptotic behavior of $\beta_{\pm}(\tau)$ is
constant, which means an effective isotropization of the solutions.
Concerning the time of  isotropization, depending on warmness
the RRG model can be closer to dust or radiation.

\section{Conclusions}
\label{Sect6}

We formulated the framework of RRG model applied to the
dynamics of anisotropy in the early epoch, where the universe
was filled by radiation and matter (baryonic and dark), which
was so hot that has the EoS which interpolates between the
radiation and pressureless matter.
For the \B1 universe away from the singularity region the
gravitational theory based on the Einstein-Hilbert action
provides an isotropization mechanism for RRG, exactly like for
both radiation and dust matter contents with isotropic EoS.
This physical situation is a subject of current interest, see,
e.g., \cite{Khalatnikov:2003ph}.
More complicated spaces may require a more complicated
gravitational theories to explain isotropization mechanism. We
believe that the simple and efficient RRG model can be useful
for describing the realistic matter contents in these complicated
cases, as it was for the rather simples \B1 universe described
above.

Another potentially interesting application of our results is
related to the cosmic perturbations in the anisotropic universe,
which is not sufficiently well explored. Since the problem is
technically complicated, it maybe very useful to have a simple
albeit realistic description of the matter contents in the early
universe in the epoch when the isotropization occurs. In this
respect the framework RRG looks perfect, since it is extremely
simple and enables one to quantify the transition from radiation
to matter epochs, exactly as it was used in the pioneer work
of Sakharov \cite{Sakharov:1966aja}.

As it was expected from the previous works on this model
\cite{FlaFlu,Fabris:2008qg,Fabris:2011am}, the RRG shows
the behaviour which is intermediate between radiation and
dust and approaches one or another  depending on the value of
warmness parameter. We have shown that this feature can be
extended to the simplest anisotropic \B1 model.

One of the natural further developments can be related to the
derivation and analysis of density and metric perturbations in
the universe filled by RRG, including the case with interaction
between RRG and radiation \cite{FAW}. One can expect that
RRG would be eventually useful as a model which helps to
explore the observables which can tell us about the dynamics
of anisotropies in the early universe.

The formalism which was developed in the present work can
be useful for the description of Bianchi I phase between the
two FLRW phases of the history of universe in the models
proposed in \cite{Comer:2011ss,Comer:2011dn}. In this
case the matter contents of the universe is supposed to be hot
and therefore the RRG can be helpful in its efficient description.

One can also use the same description of the hot or warm matter
in other approaches to anisotropy, like the recent consideration
of gravity with $R^2$ term \cite{Muller:2017nxg} or with the
sigma-model like scalar field \cite{Kamenshchik:2017ojc}, or
even in loop quantum gravity \cite{Ashtekar:2009vc}.

\section*{Appendix. Brief derivation of 
Eq.~(\ref{energymomentumtensor})}

Let us present a very brief derivation of the main expression for the 
energy-momentum tensor which was used in the main text to arrive 
at the EoS of the RRG model. More details can be found in 
\cite{Hakim2011} and also in \cite{landau1987}.

Consider a gas of free massive relativistic particles with equal
masses in the equilibrium state. Once in the comoving frame for each 
particle $T^{00}$ is the energy density, the standard arguments show 
that the energy-momentum tensor of the gas can be expressed as a 
sum over particles which are labeled by the subscript $a$,   
\beq
T^{\mu\nu}(x)
\,=\,\sum_{a}\,\int\,ds\,\,
\delta^{4}\big(x-x_{a}(s)\big)\,\,
\frac{p_{a}^{\mu}(s)\,p_{a}^{\nu}(s)}{m_{a}},
\label{em1}
\eeq
and $s$ is an integration over the proper time for individual particles. 
Using the definition of Dirac's delta function, one can rewrite 
this expression in the form
\beq
T^{\mu\nu}(x)
&=& \int d^4p\,\,\,p^{\mu}p^{\nu}\,f(x,p),
\label{em2}
\eeq
where 
\beq
f(x,p)\,=\, \sum_{a}\int ds\,
\frac{\de^4\big(p-p_{a}(s)\big)\,\de^4\big(x-x_{a}(s)\big)}{m_{a}}.
\eeq
The expression (\ref{em2}) includes an integral over four-momenta.
As far as each of the free particles satisfies a dispersion relation 
$p^2=m^2$ with $p^0\geq 0$, one can replace $d^4p$ by the 
expression 
\beq
d^3p\,dp^0\,\,\de \big(p_0^2 - {\vec p}^2 - m^2\big). 
\label{delta}
\eeq
Taking the integral over $\,p^0\,$ one has to replace the invariant 
element of integration in four dimensions $d^4p$ to the invariant 
element of integration in the space sector, $(m/p^0)d^3p$, because 
the normal vector to the $p^2=m^2$ has the same direction as 
$p^\mu$ \cite{landau1987}. Finally, using the properties of the 
delta function leads us to Eq.~(\ref{energymomentumtensor}), 
where the distribution function $f$ depends only on ${\vec p}$.  

Let us stress that the distribution function $f(x,p)$ is defined to 
be dependent on the motion of all particles. For the many-body 
system the use of the methods of Statistical Mechanics, in the case 
of a thermal equilibrium in Minkowski space leads to the distribution 
function of the J$\ddot{\rm u}$ttner model. The simplifying 
assumption of the RRG is that all particles have the same kinetic 
energy, and that is why the distribution function is chosen as a 
delta function. As we know from the previous work \cite{FlaFlu},
this approach provides an excellent approximation to complicated
EoS of the  J$\ddot{\rm u}$ttner model.
 
\section*{Acknowledgments}

Authors are very grateful to Patrick Peter for useful discussions.
S.C.R. is grateful to CAPES for supporting his Ph.D. project.
I.Sh. was partially supported by CNPq, FAPEMIG and ICTP.



\end{document}